\documentclass[floats,twocolumn,nofootinbib]{revtex4}

\usepackage{graphicx}
\usepackage{amsfonts}
\usepackage{amsmath}
\usepackage{dsfont}
\usepackage{array,hhline}

\usepackage{color}
\definecolor{Red}{rgb}{0.8,0,0}
\definecolor{Black}{rgb}{0,0,0}

\newcommand{\ZZ}{\mathbb{Z}}
\newcommand{\RR}{\mathbb{R}}
\newcommand{\CC}{\mathbb{C}}
\newcommand{\HH}{\mathbb{H}}

\newcommand{\kk}{\mathfrak{k}}
\newcommand{\calA}{\mathcal{A}}
\newcommand{\calB}{\mathcal{B}}

\newcommand{\calD}{\mathcal{D}}

\newcommand{\calH}{\mathcal{H}}
\newcommand{\calL}{\mathcal{L}}
\newcommand{\calM}{\mathcal{M}}
\newcommand{\calN}{\mathcal{N}}

\DeclareMathOperator{\UU}{U}

\DeclareMathOperator{\OO}{O}
\DeclareMathOperator{\SO}{SO}

\DeclareMathOperator{\spl}{sp}

\newcommand{\Hsp}{H_{\mathrm{s.p.}}}
\newcommand{\bound}{\mathrm{bound}}

\newcommand{\one}{\mathds{1}}
\newcommand{\otimesgr}{\mathbin{\otimes_{\mathrm{gr}}}}

\def\be{\begin{equation}}
\def\ee{\end{equation}}
\def\ba#1{\begin{array}{#1}}
\def\ea{\end{array}}
\def\bn{\begin{enumerate}}
\def\en{\end{enumerate}}

\def\H{\mathcal{H}}

\def\beq{\begin{equation}}
\def\eeq{\end{equation}}

\begin{document}

\title{Topological phases of fermions in one dimension}
\author{Lukasz Fidkowski and Alexei Kitaev}
\affiliation{California Institute of Technology, Pasadena, CA 91125, U.S.A.}

\begin{abstract}
In this paper we  show how the classification of topological phases in insulators and superconductors is changed by interactions, in the case of 1D systems.  We focus on the TR-invariant Majorana chain (BDI symmetry class). While the band classification yields an integer topological index  $k$, it is known that phases characterized by values of $k$ in the same equivalence class modulo 8 can be adiabatically transformed one to another by adding suitable interaction terms. Here we show that the eight equivalence classes are distinct and exhaustive,  and provide a physical interpretation for the interacting invariant modulo 8.  The different phases realize different Altland-Zirnbauer classes of the reduced density matrix for an entanglement bipartition into two half-chains.  We generalize these results to the classification of all one dimensional gapped phases of fermionic systems with possible anti-unitary symmetries, utilizing the algebraic framework of central extensions.  We use matrix product state methods to prove our results.

\end{abstract}

\maketitle

\section{Introduction \label{intro}}

The experimental observation of topological insulators in two and three dimensions has led to a renewed interest in the topological properties of insulators and superconductors. A classification encompassing all such band systems has been obtained by \cite{SRFL} and independently by \cite{kitaev-2009}.  Both approaches use sophisticated mathematical tools, but the topological invariants they define have a clear underlying physical interpretation: they measure the twisting of the band structure over the Brillouin zone.  The classification of \cite{SRFL} and \cite{kitaev-2009} is exhaustive, but has the drawback that it applies only to band systems, in that it defines topological invariants only in a single particle framework.  In some cases, including for example the quantized Hall conductivity or the magneto-electric susceptibility, these invariants can be interpreted as physical response functions and hence defined even in the presence of interactions, but in other cases there is no such interpretation.  In fact, \cite{fidkowski-2009} found a specific one dimensional example, the so-called Majorana chain with an unusual time reversal symmetry (TRS) squaring to $+1$, where the band classification is broken by interactions, in that some phases which are distinct in the band classification are actually connected in interacting Hamiltonian space.  Indeed, while the band classification gives an integer topological index $k$, it is at most only equivalence classes of $k$ modulo $8$ that define distinct interacting phases: ${\mathbb Z}$ is broken down to ${\mathbb Z}_8$.  The TR-invariant Majorana chain example thus highlights a deficiency of the band classification.

In this paper we re-examine the TR-invariant Majorana chain, and provide an interacting interpretation for the eight different phases.  Before delving into the construction, let us first provide some intuition.  The band ${\mathbb Z}$ invariant can be interpreted as the number of gapless Majorana modes localized at an endpoint of the chain.  To see how it is broken, we proceed by analogy: consider an isotropic, TR-invariant spin-$1$ chain, which has two phases - a trivial phase, and a topological ``Haldane'' phase. The spin chain can also have gapless spins localized at an endpoint, but different values of this edge spin do not correspond to different phases: rather, it is only whether the edge spin is integral or half-integral that determines the phase (trivial or Haldane) of the system.  Stated more abstractly, it is the symmetry class of the edge spin -  ``real'' ($T^2=1$) or ``quaternionic'' ($T^2 = -1$)  - that determines the phase \cite{ari2} (see also \cite{HKH2, HKH1}).  The Majorana chain Hamiltonian respects fermionic parity $(-1)^F$ in addition to $T$, modifying the set of possible symmetry classes of the `edge spin'. In fact, there are now eight phases, and they are in one to one correspondence with $k\bmod 8$. 

In our construction we will make use of the notion of an entanglement spectrum: rather than looking at systems with physical edges, the most convenient framework for us will be to simulate edges with entanglement bipartitions.  The advantage of this approach is that it is manifestly independent of the details of an edge Hamiltonian and does not break any extra symmetries.  Indeed, the study of topological phases via their entanglement structure has a rich history \cite{kitaev-preskill, levin-wen, haldane-li, bernevig, ari1, ari2, fidkowski}.  Of particular interest to us is the connection between the entanglement spectrum and edge mode spectrum found in \cite{haldane-li} for the fractional quantum Hall effect and in \cite{ari1, fidkowski} for band topological insulators and superconductors.  This connection suggests that the entanglement spectrum behaves like the edge mode spectrum, and in particular that we should study the structure of representations of the generic symmetries (such as $T$ and $(-1)^F$) on it.  This proves to be a fruitful approach: already in \cite{ari2} it was shown that the trivial and Haldane phase in the spin-$1$ Heisenberg chain are distinguished by $T^2=\pm 1$ on the entanglement spectrum.  Here we generalize the construction to an arbitrary fermionic chain with  a real  time reversal symmetry, i.e. the TR-invariant Majorana chain.  The final result is rather elegant: the phases of the TR-invariant Majorana chain are in one to one correspondence with eight of the ten Altland-Zirnbauer (AZ) classes \cite{AZ,Zirnbauer,HHZ} (used also in the scheme of \cite{SRFL,RSFL1}), and the signature of the phase one is in is the AZ symmetry class of the reduced density matrix of half of an infinite chain.

As in \cite{ari2}, the key technical tool that allows rigorous arguments is that of matrix product states (MPS).  The powerful entropy scaling bound \cite{hastings} for gapped one dimensional systems allows us to approximate the ground state of any such infinite chain by a MPS of fixed bond size, depending only on the gap and desired accuracy of approximation.  The entanglement spectrum for a MPS is simply the bond Hilbert space, and known results \cite{wolf} classify the possible representations of the global symmetries on the bond Hilbert space - they are so-called projective representations, and for the TR-invariant Majorana chain they can be used to construct invariants that are in one to one correspondence with the Altland-Zirnbauer symmetry classes.  Our arbitrarily accurate MPS approximations then extend this result to general gapped chains.

The remainder of the paper is organized as follows.  In section (\ref{bc}) we establish notation, define the TR-invariant Majorana chain, and review its band classification.  In section (\ref{mps}) we review necessary facts about matrix product states, and show how symmetries lead to projective invariants.  In section (\ref{ii}) we use these results to classify the phases of the interacting TR-invariant Majorana chain.  In section (\ref{mc}) we extend our scheme to the general classification of the phases of one dimensional gapped systems with both unitary and anti-unitary symmetries.  The classification uses the algebraic notion of central extensions, which is a precise way of defining projective invariants.  For completeness, in section (\ref{ex}) we compute the invariants for flat-band models representative of the $8$ phases of the TR-invariant Majorana chain, showing that all possibilities are realized, and relate them to the value of the topological index $k$ modulo $8$.  We conclude with a discussion of related matters and future directions in section (\ref{d}).  In the appendices we discuss in more depth the mathematical structure common to Altland-Zirnbauer theory and the TR-invariant Majorana chain, and give a short review of semisimple algebras.

As we were writing this article, we learned of the simultaneous independent work of \cite{turner3}, whose results agree with ours.  Furthermore, after the completion of this work, we noticed \cite{wen-new}, which also points to a similar classification of gapped phases of one dimensional chains.

\section{Preliminaries \label{bc}}

We start with a second quantized Hamiltonian $H$ in the creation and annihilation operators $a^\dag_j, a_j$ of spinless fermions, where $j$ indexes sites of a chain.  We assume $H$ is gapped and includes only short range bounded strength interactions.  We allow possible pairing terms, e.g. $a_j^{\dag}a_k^{\dag}$. In this basis, the time reversal symmetry $T$ acts as complex conjugation of the wavefunction, while the creation and annihilation operators are real: \begin{eqnarray} T \, a_j \, T^{-1} &=& a_j, \nonumber \\ T \, a^\dag_j \, T^{-1} &=& a^\dag_j. \end{eqnarray} Let us now restrict to quadratic Hamiltonians and review the band classification of the phases of the Majorana chain.  It is convenient to introduce the Hermitian ``Majorana'' operators \begin{eqnarray} c_{2j-1} &=& -i\, (a_j - a^\dag_j) \nonumber \\ c_{2j} &=& a_j + a^\dag_j \end{eqnarray}  Any Hermitian quadratic Hamiltonian $H$ can then be written in the form
\begin{equation}\label{quadH}
H = \frac{i}{4} \, \sum_{l,m} A_{lm} \, c_l \, c_m,
\end{equation}
where $A$ is a real anti-symmetric matrix.

\begin{figure}[htp]
\includegraphics[width=6.5cm]{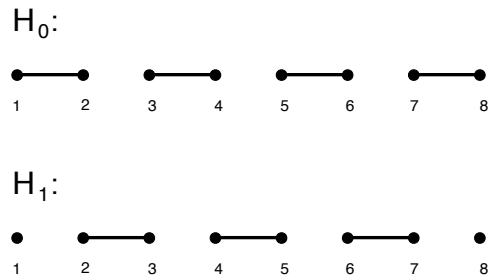}
\caption{Schematic representation of the Hamiltonians $H_0$ and $H_1$.  The dots denote Majorana fermions, and the edges the quadratic couplings in the Hamiltonian. \label{zfig1}}
\end{figure}

First, assume that we do not impose $T$: there are then two distinct phases of the Majorana chain \cite{kitaev-2009}: the trivial phase
\begin{eqnarray}
\label{h0} H_0 &=& -\frac{i}{2} \sum_{j=1}^N c_{2j-1} \, c_{2j} \\
&=& \sum_j \left(a^\dag_j \, a_j - \frac{1}{2}\right)
\nonumber
\end{eqnarray}
where all sites are decoupled and unoccupied in the ground state, and a non-trivial phase 
\begin{eqnarray}
\label{h1} H_1 &=& -\frac{i}{2} \sum_{j=1}^{N-1} c_{2j} \, c_{2j+1} \\
&=& \frac{1}{2} \sum_j
\left(- a^\dag_j\,a_{j+1} - a^\dag_{j+1}\,a_j
+ a_j^\dag\,a_{j+1}^\dag + a_{j+1} \, a_j \right).
\nonumber
\end{eqnarray}
These are illustrated in figure \ref{zfig1}.  The defining characteristic of the non-trivial phase is a two-fold ground state degeneracy: the edge ``dangling'' Majorana operators $c_1$ and $c_{2N}$ can be paired up into a physical fermion mode that does not cost any energy to occupy.


\begin{figure}[htp]
\includegraphics[width=6.5cm]{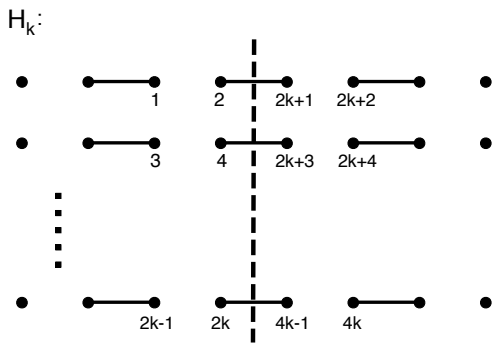}
\caption{Schematic representation of the Hamiltonian $H_k$.  The dashed vertical line can represent a physical cut or an entanglement bipartition.  Note that on the left side of the cut we have $k$ unpaired Majorana fermions, $c_2, \ldots, c_{2k}$. \label{zfig3}}
\end{figure}

Now impose $T$.  There are then infinitely many phases, indexed by an integer $k$ \cite{kitaev-2009}.  We can interpret these phases in terms of dangling Majorana modes as well.  We have infinitely many phases, instead of just two, because $T$ symmetry restricts the allowed interactions, preventing us from gapping out dangling Majorana modes in pairs.  For example, consider the Hamiltonian $H_k$ illustrated in figure \ref{zfig3}, and take $k=2$.  If we cut the chain along the dashed line, the unpaired Majorana modes $c_2$ and $c_4$ on the left side cannot be gapped out, because the interaction $i\,c_2c_4$ is not $T$ invariant by virtue of $2$ and $4$ having the same parity.  Thus $H_2$ represents a new $T$-protected phase, distinct from the trivial one.  One can have phases with any number $k$ of such dangling Majorana modes whose indices have the same parity, leading to an integer band topological index $k$ - the corresponding Hamiltonians $H_k$ are represented in diagram form in figure \ref{zfig3}: we think of $H_k$ as the Hamiltonian of $k$ parallel Majorana chains, which can be viewed as a one dimensional chain when one chooses an appropriate ordering of the Majorana sites.  It is only the equivalence class of $k$ modulo $8$, however, that is well defined in the interacting setting, as we demonstrate in the rest of the paper.

In order to define the interacting invariants, we will need to use matrix product states (MPS).  To do this, we must first do a Jordan-Wigner transform to a bosonic spin chain.  This is a general procedure that works equally well for interacting systems: we define
\begin{eqnarray}
\sigma^x_j &=& \left(a_j + a^\dag_j \right) \,
\prod_{k<j}\left(1-2a^\dag_k a_k\right), \nonumber\\
\sigma^y_j &=& -i\left(a_j - a^\dag_j \right) \,
\prod_{k<j}\left(1-2a^\dag_k a_k\right), \nonumber \\
\sigma^z_j &=& 1- 2 a^\dag_j a_j.
\end{eqnarray}
The ${\mathbb Z}_2$ fermionic parity is given by the operator
\begin{equation}
\label{peq} P = \prod_{j} \left(i\,c_{2_j-1}c_{2j}\right)
= \prod_j\sigma^z_j.
\end{equation}
In general, the Jordan-Wigner transforms of the Hamiltonians $H_k$ have ground states which spontaneously break $P$ if and only if $k$ is odd.  This can be explicitly seen by expressing the exact ground states in bosonic variables, resulting in so-called cluster states \cite{Gross}, but is easier to demonstrate via explicit calculation for $H_0$ (\ref{h0}) and $H_1$ (\ref{h1}).  The Jordan-Wigner transform of $H_0$ is  \begin{equation} \tilde{H}_0 = -\frac{1}{2} \sum_j \sigma_j^z. \end{equation}  whose ground state of all spins pointing up is an eigenstate of $P$.  The Jordan-Wigner transform of (\ref{h1}), on the other hand, is the Ising Hamiltonian \begin{equation} \tilde{H}_1 = -\frac{1}{2}\sum_j\sigma^y_j\sigma^y_{j+1}, \end{equation}  whose two ground states spontaneously break $P$.  We stress that this symmetry breaking occurs only in the bosonic spin chain, and is a result of the non-local nature of the Jordan-Wigner transformation. $P$ is never broken in the fermionic Majorana chain.  In the fermionic language, the dimers formed by $c_{2j}$ and $c_{2j-1}$ are invariant under the action of $P$, and therefore $P$ may be replaced by
\begin{equation}
\hat{P}=i\,c_{1}c_{2N}.
\end{equation}
Thus, the even and the odd superpositions of the bosonic ground states correspond to the ``empty'' and the ``occupied'' states of the pair of dangling Majorana modes.

\section{Interacting Invariants via Matrix Product States \label{mps}}

Let us now review some facts about MPS which we will need in the following sections \cite{vidal, wolf, wolf2, ari2}. We assume translational invariance for convenience, though we believe our results to be valid even without it. An MPS on a one dimensional chain is a quantum state $|\Psi\rangle$ whose Schmidt decomposition across any cut, say, between sites $n-1$ and $n$, has bounded rank $\alpha_{\mathrm{max}}$:
\begin{equation} \label{mpses}
|\Psi \rangle = \sum_{\alpha=1}^{\alpha_{\mathrm{max}}} \lambda_{\alpha} \,
|\Psi_{n-1}^{\alpha L} \rangle \otimes |\Psi_{n}^{\alpha R} \rangle,\qquad
\lambda_{\alpha}> 0.
\end{equation}
Had we performed the cut between sites $n$ and $n+1$, we would have obtained the same Schmidt decomposition, but with eigenvectors $|\Psi_{n}^{\alpha L} \rangle$ and $|\Psi_{n+1}^{\alpha R} \rangle$.  We can write one set of Schmidt vectors in terms of the other; for example
\begin{equation}
|\Psi_{n}^{\beta L} \rangle
= \sum_\alpha A_m^{\alpha \beta} |\Psi_{n-1}^{\alpha L} \rangle
\otimes |m\rangle.
\end{equation}
In a more abstract language, $A$ is a linear map of type $\calH_L\to\calH_L\otimes\calH_{\mathrm{spin}}$, where $\calH_L$ is spanned by the left Schmidt vectors. We refer to $\calH_L$ as the \emph{bond Hilbert space}, or \emph{entanglement Hilbert space}. Thus, $A_m:\,\calH_L\to\calH_L$ for each $m$.

To find the inherent constraints on the matrices $A_m$, let
\begin{equation}\label{trop}
E: X \mapsto \sum_m A_m^\dag X A_m,
\end{equation}
and
\begin{equation}
E^*: X \mapsto \sum_m A_m X A_m^\dag.
\end{equation}
Then the properties that the two sets of Schmidt vectors are orthonormal and the corresponding numbers $\lambda_\alpha$ are the same imply that
\begin{equation}\label{mpsrel}
E(\one)=\one,\qquad E^*(\Lambda^2)=\Lambda^2.
\end{equation}
In other words, $\one=(\delta_{\alpha\alpha'})$ and $\Lambda^2=(\lambda_{\beta}^{2}\delta_{\beta\beta'})$ are a left and right eigenvector associated with the $\nu=1$ eigenvalue of the transfer matrix
\begin{equation}\label{tm}
E_{\alpha\alpha';\beta\beta'} = \sum_{m} A^{\alpha\beta}_{m} \bigl(A^{\alpha'\beta'}_{m}\bigr)^*.
\end{equation}
If this eigenvalue is simple and there are no other eigenvalues of magnitude $1$, the MPS (or rather, the set of matrices defining it) is called \emph{simple}, or ergodic. It is easy to see that all eigenvalues satisfy $|\nu|\le 1$ (for example, because $E$ does not increase the operator norm).

Proceeding iteratively, we can write
\begin{equation}
| \Psi \rangle =
\Bigl(\cdots A^{\alpha\beta}_{m_{k-1}} A^{\beta\gamma}_{m_k}
 A^{\gamma\delta}_{m_{k+1}}\!\cdots \Bigr)
|\ldots m_{k-1} m_k m_{k+1}\ldots \rangle,
\end{equation}
where we implicitly sum over repeated indices. The simplicity condition mentioned above implies that the infinite state $|\Psi\rangle$ is pure. When we approximate ground states of the Jordan-Wigner transformed TR-invariant Majorana chain, this occurs for even topological index $k$.  When $k$ is odd, there are two degenerate ground states, i.e. the ground state density matrix has rank $2$.  It can be approximated with a mixed MPS whose transfer matrix has two eigenvalue $\nu = 1$ eigenvectors. 

Since one may feel uncomfortable with infinite states, let us calculate the density matrix of a finite segment. The equations will look more symmetric if we define the matrices $\Gamma_m$ by
\begin{equation}
A_m =\Lambda\Gamma_m.
\end{equation}
Then,
\begin{gather}
\rho_{[0,n]}=\sum_{\alpha,\beta}\lambda_{\alpha}^2\lambda_{\beta}^2\,
\bigl|\Psi_{[0,n]}^{\alpha\beta}\bigr\rangle\,
\bigl\langle\Psi_{[0,n]}^{\alpha\beta}\bigr|,\\[3pt]
\bigl|\Psi_{[0,n]}^{\alpha\beta}\bigr\rangle=\!\sum_{m_1,\ldots,m_n}\!
\bigl(\Gamma_{m_1}\Lambda\cdots\Lambda\Gamma_{m_n}\bigr)^{\alpha\beta}
|m_{1}\ldots m_{n}\rangle.
\end{gather}
If the MPS data are simple, then the states $\bigl|\Psi_{[0,n]}^{\alpha\beta}\bigr\rangle$ become orthonormal in the $n\to\infty$ limit, indicating that the segment endpoints decouple.

For our purposes, all nontrivial properties of MPS's can be derived from the following result \cite{wolf, wolf2}.\medskip

\noindent\textbf{Theorem.} Two sets of simple MPS data, $(\Gamma_m,\Lambda)$ and  $(\Gamma_m',\Lambda')$ define the same physical state (i.e. $\rho_{[0,n]}=\rho_{[0,n]}'$ for all $n$) if and only if there is some unitary matrix $U$ and number $\theta$ such that
\begin{equation}\label{Utheta}
\Gamma_m'=e^{i\theta}U^{-1}\Gamma_{m}U,\qquad \Lambda'=U^{-1}\Lambda U.
\end{equation}
The factor $e^{i\theta}$ is unique, and $U$ is unique up to a phase.\medskip

The uniqueness can be shown pretty easily. Let us suppose that eq.~(\ref{Utheta}) has two solutions, $(U_1,\theta_1)$ and $(U_2,\theta_2)$, and let $V=U_1U_2^{-1}$. Then $V$ commutes with $\Lambda$ and $\Gamma_m$ (up to a phase), and hence with $A_m=\Lambda\Gamma_m$. Applying (\ref{trop}) and (\ref{mpsrel}), we find that \[
E(V)=e^{i(\theta_1-\theta_2)}\,E(\one)\,V=e^{i(\theta_1-\theta_2)}V
\]
so that $V$ is an eigenvector associated with a magnitude $1$ eigenvalue of the transfer matrix. Due to the simplicity assumption, $1$ is the only such eigenvalue, and the eigenvector is proportional to the identity matrix. For the proof of existence, see \cite{wolf, wolf2}.

Suppose now that we have a simple MPS and a finite global internal symmetry group $G$, whose elements $g$ act in the physical spin space by a matrix representation $(g_{mn})$. The corresponding matrices $\Gamma_m$ are transformed into $\Gamma_m'=\sum_n g_{mn} \Gamma_n$. In the case of anti-unitary $g$ the relevant operator is the unitary matrix $(g_{mn})$ composed with complex conjugation, i.e. $\Gamma_m'=\sum_n g_{mn} \Gamma_n^{*}$. Since $\Gamma_m$ and $\Gamma_m'$ represent the same physical state, the above theorem gives an induced action of $G$ in the bond Hilbert space \cite{ari2, wolf}.  That is, for each $g$ there is a corresponding operator $\hat{g}$, whose explicit form in some basis is
\begin{equation} \hat{g} = \left\{ \begin{array}{ll}
U_g, &\text{if $g$ is unitary},\\[2pt]
U_g K, &\text{if $g$ is anti-unitary},
\end{array}\right.
\end{equation}
where $U_g$ is a unitary matrix and $K$ is complex conjugation. Such $\hat{g}$ is defined up to a phase, commutes with $\Lambda$, and satisfies
\begin{equation} \label{up}
\sum_n g_{mn} \left({\hat{g}}\,\Gamma_n{\hat{g}}^{-1}\right)
= e^{i \theta_g}\Gamma_m
\end{equation}
for some $\theta_g$. Because $\hat{g}\kern1pt\hat{h}$ and $\widehat{gh}$ are associated with the same transformation $gh$ of physical spins, the uniqueness part of the above theorem guarantees that they are equal up to phase, showing that the operators $\hat{g}$ form a projective representation of $G$ on the bond Hilbert space.

This analysis is applicable to an arbitrary gapped spin chain whose Hamiltonian $H$ is invariant under $G$ and has a unique ground state. Indeed, the ground state can be approximated arbitrarily well with MPS \cite{hastings}, allowing us to define a projective representation of $G$ on its entanglement Hilbert space as well: $\hat{g}:\calH_L\rightarrow\calH_L$.  A group element $g$ acts on operators localized on the left half of the chain by $X\mapsto\sigma_{g}(X)=\hat{g}X\hat{g}^{-1}$. We will use this action to construct invariants which characterize the gapped phases of $H$. In the mathematical literature such invariants go by the name of extensions of $G$ with $\UU(1)$, and (for a finite $G$) are characterized by the group cohomology with $\UU(1)$ coefficients.  Informally, they encode in an invariant way the extent to which the phase ambiguities inherent in the definition of $\hat{g}$ are incompatible with the original group structure of $G$. But different projective representations corresponding to the same extension are compatible with each other. For example, the entanglement Hilbert space of a spin-$1$ chain may include sectors corresponding to different half-integral values of the the bond spin, e.g.\ $1/2$ and $3/2$. The matrices $\Gamma_n$ can mix them according to the fusion rule $[1/2]\times[1]=[1/2]+[3/2]$. On the other hand, integral and half-integral values of the bond spin cannot coexist without breaking the ergodicity.

For finite $G$, these invariants are discrete, and using MPS approximations can be shown to vary continuously as one moves within the same gapped phase, implying they are in fact constant in that phase. Hence they are many-body invariants characterizing interacting gapped phases of the spin chain.  We will discuss their general mathematical structure, for both unitary and anti-unitary symmetries, in section (\ref{mc}); first, however, we explicitly construct them for the relatively simple case of the TR-invariant Majorana chain. 

\section{TR-invariant Majorana chain \label{ii}}

In the TR-invariant Majorana chain, the symmetry group $G$ is generated by the fermionic parity $P$ and time reversal $T$.  These satisfy the commutation relations $P^2 = T^2 = 1$ and $PT=TP$.  What invariants can we construct for the projective action of $G$ on the entanglement Hilbert space?  Let us begin by considering even topological index $k$, where the ground state is a simple MPS.  We have  two non-trivial operators, $\hat{P}$ and $\hat{T}$, as well as their product $\hat{P}\hat{T}$.  The projective form of the relation $P^2=1$ is $\hat{P}^2=e^{i \phi} \one$ for some $\phi$; however, $\phi$ can be removed by a phase shift of $\hat{P}$, so that we can assume $\hat{P}^2 = \one$.\,  $T$ and $PT$ also square to $1$, but because they are anti-unitary, the corresponding projective operators have to square to the identity up to sign: $\hat{T}^2 = \pm \one$,\:  $(\hat{P}\hat{T})^2 = \pm \one$.  These signs cannot be removed by any phase rotations, and together define four invariants for the projective representations, distinguishing four even $k$ phases of the TR-invariant Majorana chain.  To reformulate slightly, the phases are distinguished by whether $\hat{T}$ is  real or quaternionic (column $D$ in table~\ref{table1}), and whether it commutes or anti-commutes with the projective fermionic parity operator $\hat{P}$ (the invariant $a=\pm1$). These invariants as functions of $k$ will be calculated in section~\ref{ex}.  Now, the reduced density matrix of the left half of the chain, $\Lambda^2$, contains both parity-even and parity-odd sectors, and $\hat{T}$ commuting or anti-commuting with $\hat{P}$ means that $\hat{T}$ preserves or exchanges these sectors, respectively.  Thus, the four different phases are distinguished by whether the anti-unitary symmetry $\hat{T}$ is  real or quaternionic, and whether it is to be interpreted as ``time reversal symmetry'' $\hat{T}_{+}$ or ``particle-hole symmetry'' $\hat{T}_{-}$. This determines the orthogonal, symplectic, and C and D BdG Altland-Zirnbauer classes, see table~\ref{table1}. (The translation to the AZ language is formalized in appendix~\ref{AZW}.)

For odd topological index $k$ one can perform a similar analysis, but it is complicated by the fact that there are now two degenerate ground states, $|\Psi' \rangle$ and $|\Psi'' \rangle$.  We can approximate $|\Psi' \rangle$ by an MPS $(\Gamma'_m, \Lambda')$.  Then because $|\Psi'' \rangle$ is related to $|\Psi' \rangle$ by the $\mathbb{Z}_2$ symmetry $P$ (\ref{peq}), a good MPS approximation to it is given by the MPS  \begin{eqnarray} \label{gammaprime} \left(\Gamma''_m, \Lambda''\right) &=& \left(\sum_n P_{mn} \Gamma'_n,\: \Lambda'\right) \nonumber \\ &=& \bigl((-1)^m\Gamma'_m,\, \Lambda'\bigr),\end{eqnarray} where $m=0$ and $m=1$ refer to the ``up'' and ``down'' spin states, respectively. According to this definition, the associated operator $\hat{P}'$ on the bond Hilbert space is trivial, since the relation between $\Gamma'$ and $\Gamma''$ is entirely due to the action of $P$ on the physical spin. (Later we will define a nontrivial $\hat{P}$ acting in a larger space.)

Consider now the action of time reversal symmetry. The states $|\Psi' \rangle$ and $|\Psi'' \rangle$ obey the clustering condition on correlation functions in the thermodynamic limit, and the action of $T$ preserves this clustering condition.  Thus $T$ must either fix  $|\Psi' \rangle$ and $|\Psi'' \rangle$ individually,  or exchange them.  Let us analyze each of these cases in turn.  

If $T$ fixes the states, then the simple MPS $(\Gamma'_m, \Lambda')$ has a projective representation $\hat{T}'$ of $T$ on the bond Hilbert space, with the two possibilities: $(\hat{T}')^2 = \pm \one$.  By the definition (\ref{gammaprime}) of $(\Gamma''_m, \Lambda'')$ we see that these MPS data are preserved if the symmetry $T=PTP^{-1}$ acting on the state $| \Psi'' \rangle$ is accomplished by the operator $\hat{T}'' = \hat{P}' \hat{T'}(\hat{P}')^{-1}=\hat{T}'$. In the case that $T$ exchanges $|\Psi' \rangle$ with $|\Psi'' \rangle$, we may apply the above argument to the symmetry $PT$, which fixed each state.

The previous arguments are only applicable to ``unbroken'' symmetries. To represent other symmetries (e.g.\ $T$ in the second case, where it exchanges $|\Psi'\rangle$ and $|\Psi''\rangle$\,), we double the bond Hilbert space and consider the nonsimple MPS that corresponds to the total (mixed) ground state. It is the direct sum,
\begin{equation}
\Gamma_m = \begin{pmatrix} \Gamma'_m&0\\0&\Gamma''_m \end{pmatrix},\qquad \Lambda = \begin{pmatrix} \Lambda'&0\\0&\Lambda'' \end{pmatrix}, \end{equation}
with the symmetry $\hat{P}$ acting as
\begin{equation}\label{formofp}
\hat{P} = \sigma^x\otimes\one =
\begin{pmatrix} 0 & \one \\ \one & 0 \end{pmatrix}.
\end{equation}
Furthermore, we can define an additional operator
\begin{equation}\label{formofz}
\hat{Z} = \sigma^z\otimes\one =
\begin{pmatrix} \one & 0 \\ 0 & -\one \end{pmatrix},
\end{equation} which distinguishes the sectors of the Hilbert space corresponding to $|\Psi'\rangle$, $|\Psi''\rangle$. We have $\hat{Z}\hat{P}=-\hat{P}\hat{Z}$. In this representation, the symmetry $T$ acts as
\begin{equation}
\hat{T} = \begin{pmatrix} \hat{T}' & 0 \\ 0 & \hat{T}' \end{pmatrix}
\quad\text{or}\quad
\hat{T} = \begin{pmatrix} 0 & \hat{T}' \\ \hat{T}' & 0 \end{pmatrix}
\end{equation}
if $\hat{T}$ fixes or exchanges the two states, respectively. We see that $\hat{P}\hat{Z}=a\hat{Z}\hat{P}$, where $a$ is equal to $1$ in the first case and to $-1$ in the second case.

To translate to the AZ language, we define the ``time-reversal symmetry'' by $\hat{T}_{+}=\hat{T}$ (because $\hat{T}$ always commutes with $\hat{P}$). The ``sublattice symmetry'' $\hat{Z}$ should be equal to $\hat{T}_{+}\hat{T}_{-}$, therefore $\hat{T}_{-}=\hat{T}^{-1}\hat{Z}$. Thus we have obtained the chiral orthogonal and chiral symplectic symmetry classes for $a=1$, and the CI and DIII BdG classes for $a=-1$ (see table~\ref{table1}). 

\section{General Classification \label{mc}}

In this section we consider an arbitrary gapped one dimensional system that consists of spins or itinerant fermions and has an unbroken symmetry group involving both unitary and anti-unitary elements. (Let us reiterate that the fermionic parity $P$ is also always unbroken - the ${\mathbb Z}_2$ breaking discussed in the previous sections was in the Jordan-Wigner transformed bosonic spin chain, which is related in a non-local way to the fermionic chain.) The TR-invariant Majorana chain is in some sense the simplest non-trivial example, in that the Hamiltonian is invariant under both the unitary fermionic parity $P$ and anti-unitary time reversal $T$.  Having shown how to rigorously define the invariants for the TR-invariant Majorana chain with entanglement bipartitions, we shift our point of view slightly in this section, and discuss systems with physical edges - that is, finite segments. We will be concerned with low energy degrees of freedom localized at the endpoints of the segments.  Although all the properties we need can be proven by translating the statements to the language of entanglement bipartitions and using MPS as in the previous sections, here we will be content with simply clearly stating the necessary properties.  These will have to do with the form of the low energy sub-spaces, and the space of operators acting on them.

We begin by considering a bosonic spin system defined on a one dimensional lattice, with a Hamiltonian $H$ that is gapped in the bulk and is invariant under the simultaneous action of a symmetry group $G$ on all sites. This action may be projective, which allows us, for example, to treat a (dimerized) spin-$1/2$ Heisenberg chain as a system with $\SO(3)$ symmetry. In any case, there is a genuine action of $G$ on physical observables, and that is sufficient for our purposes. Now we imagine a finite chain, with left and right endpoints.  The Hilbert space $\calL_{\bound}$ of low energy boundary states is then assumed to decompose as
\begin{equation}
\calL_{\bound} = \calL_{l} \otimes \calL_{r}.
\end{equation}
We find it more useful to consider the algebras of linear operators defined on these Hilbert spaces:
\begin{equation}
\CC\bigl(\calL_{\bound}\bigr)
= \CC\bigl(\calL_{l}\bigr) \otimes \CC\bigl(\calL_{r}\bigr).
\end{equation}
The group $G$ has a well defined action by automorphisms on such operators.

Let us specialize to the right endpoint, and denote the action of $g \in G$ on $\calA_{r}=\CC(\calL_{r})$ by $X\mapsto \sigma_g(X)$. We first consider only unitary symmetries. The map $\sigma_g$ being an automorphism, preserves scalars and satisfies
\begin{eqnarray}
\sigma_g(X+Y)&=&\sigma_g(X)+\sigma_g(Y),\nonumber\\
\sigma_g(XY)&=&\sigma_g(X)\,\sigma_g(Y),\nonumber\\
\sigma_g(X^{\dag})&=&\sigma_g(X)^{\dag}.\label{alg_auto}
\end{eqnarray}
Because $\calA_{r}$ is a matrix algebra, and hence simple, every automorphism must be of the form $\sigma_g(X) = U_g \, X \, U_g^{-1}$ for some unitary $U_g$, which is well defined up to a phase. To accommodate future generalizations, let us use an alternative notation: $\hat{g}=U_g$. We fix the phase for each $g$ in an arbitrary way, but also consider an invariant object, the group ${\tilde G}$ that consists of operators of the form $e^{i\phi}\,\hat{g}$. Thus we have an exact sequence
\begin{equation}
1 \rightarrow \UU(1) \rightarrow {\tilde G} \rightarrow G \rightarrow 1. \end{equation}
The isomorphism classes of such exact sequences are known as extensions of $G$ with $\UU(1)$. For example, the group $G=\SO(3)$ has two extensions: $\tilde{G}\cong\SO(3)\times\UU(1)$ or $\tilde{G}\cong\UU(2)$. Representations of the two variants of $\tilde{G}$ correspond to integral and half-integral spin, respectively, where we require that the $\UU(1)$ subgroup act by the multiplication by scalars. The first case is realized when the spin-$1/2$ chain is cut between dimers, whereas the second case describes the chain cut across a dimer. Another interpretation is the trivial and the Haldane phase in a spin-$1$ chain. As shown in Refs.~\cite{ari1,ari2}, the two phases remain distinct when the symmetry is reduced to the dihedral subgroup $D_2\subset\SO(3)$. Indeed, $D_2\cong\ZZ_2\times\ZZ_2=\{I,X,Y,XY\}$ also has two extensions. The nontrivial one is given by, e.g., $U_{I}=1$,\, $U_{X}=\sigma^x$,\, $U_{Y}=\sigma^y$,\, $U_{XY}=\sigma^z$.

In the following discussion we take $G$ to be finite. Extensions of $G$ with $\UU(1)$ can be described explicitly using 2-cochains, i.e., $\UU(1)$-valued functions $C(g,h)$. These are defined by
\begin{equation}
\hat{g}\kern1pt\hat{h} = C(g,h) \,\widehat{gh}.
\end{equation}
Associativity gives the cocycle constraint,
\begin{equation}
C(g,h) \,C(fg,h)^{-1} \,C(f,gh) \,C(f,g)^{-1} = 1,
\end{equation}
while redefinition of the $U_g$ by a $\UU(1)$ phase leads to the gauge symmetry $C \sim \Delta\cdot C$, where the right-hand side is the product of $C(g,h)$ and $\Delta (g,h) = B(h)\,B(gh)^{-1}\,B(g)$ for some function (1-cochain) $B$.  Gauge equivalence classes of functions $C(g,h)$ satisfying the associativity constraint constitute the cohomology group $H^2(G,\UU(1))$. As we have seen, they are in one to one correspondence with $\UU(1)$ extensions of $G$.

With a slight modification to this framework we can incorporate anti-unitary symmetries. The setup now is that $G$ is equipped with a homomorphism $\alpha: G \rightarrow {\mathbb Z}_2$, indicating whether a given element $g$ is unitary or anti-unitary. The action of $G$ on scalars is generally nontrivial,
\begin{equation}\label{act_scal}
\sigma_g(c)=\left\{\begin{array}{ll}
c, &\text{if } \alpha(g)=+1,\\[2pt]
c^*, &\text{if } \alpha(g)=-1,
\end{array}\right.
\end{equation}
while equations~(\ref{alg_auto}) still hold. Each automorphism $\sigma_g$ is still represented by an operator $\hat{g}$ acting in the space of boundary states $\calL_{r}$.  If we fix a basis, we can write $\hat{g} = U_g$ or $\hat{g} = U_g \, K$, where $U_g$ is a unitary matrix and $K$ is complex conjugation.  The gapped phases are still classified by $\UU(1)$ extensions, or elements of $H^2(G,\UU(1))$, where the action~(\ref{act_scal}) of $G$ on $\UU(1)$ is assumed. This amounts to a simple modification of the cocycle condition,
\begin{equation}
\sigma_f (C(g,h))\, C(fg, h)^{-1} \,C(f,gh) \,C(f,g)^{-1} = 1
\end{equation}
and the equivalence relation: $C\sim\Delta\cdot C$, where
\begin{equation}
\Delta(g,h) = \sigma_g(B(h))\,B(gh)^{-1}\,B(g).
\end{equation}
The last equation corresponds to the gauge transformation $\hat{g} \mapsto B(g)\, \hat{g}$ (the multiplication order matters). One example, $G=\{I,T\}$ with $T$ anti-unitary, was mentioned in the introduction.

The incorporation of fermions is slightly more involved. We now have a special unitary symmetry, the fermionic parity $P \in G$ which is involutory ($P^2=I$) and central in $G$. Operator algebras are $\ZZ_2$-graded, $\calA=\calA^{(0)}\oplus\calA^{(1)}$. Namely, an operator $X$ is even (odd) if it preserves (resp., changes) the number of fermions modulo~2, i.e., 
\begin{equation}
X\in\calA^{(x)}\quad \text{iff\: } \sigma_{P}(X)=(-1)^{x}X\qquad (x=0,1).
\end{equation}
At the elementary level, such operators are sums of even (odd) products of $a_j,a^{\dag}_j$. Note that all automorphisms $\sigma_g$ respect the grading. Indeed, if $X\in\calA^{(x)}$, then
\[
\sigma_{P}\bigl(\sigma_g(X)\bigr)=\sigma_{g}\bigl(\sigma_P(X)\bigr)
=(-1)^{x}\sigma_{g}(X),
\]
implying that $\sigma_{g}(X)\in\calA^{(x)}$.

The algebras $\calA_{l}$, $\calA_{r}$ of operators acting at the ends of the fermionic chain are simple in the graded sense. Instead of defining this notion abstractly, we will describe the structure of such algebras and give some physical interpretation (for the general theory of simple $\ZZ_2$-graded algebras over an arbitrary field, see Ref.~\cite{Wall}).  Note that $\calA_{r}$ consists of operators that act near the right endpoint \emph{in the actual system} rather then its Jordan-Wigner transformed version. The MPS interpretation is slightly different; let us only mention that the operator  $\hat{Z}$ (see eq.~(\ref{formofz})) belongs to $\calA_{r}$, but $\hat{P}$ artificially extends that algebra. 

\begin{figure}[htp]
\includegraphics[width=6.5cm]{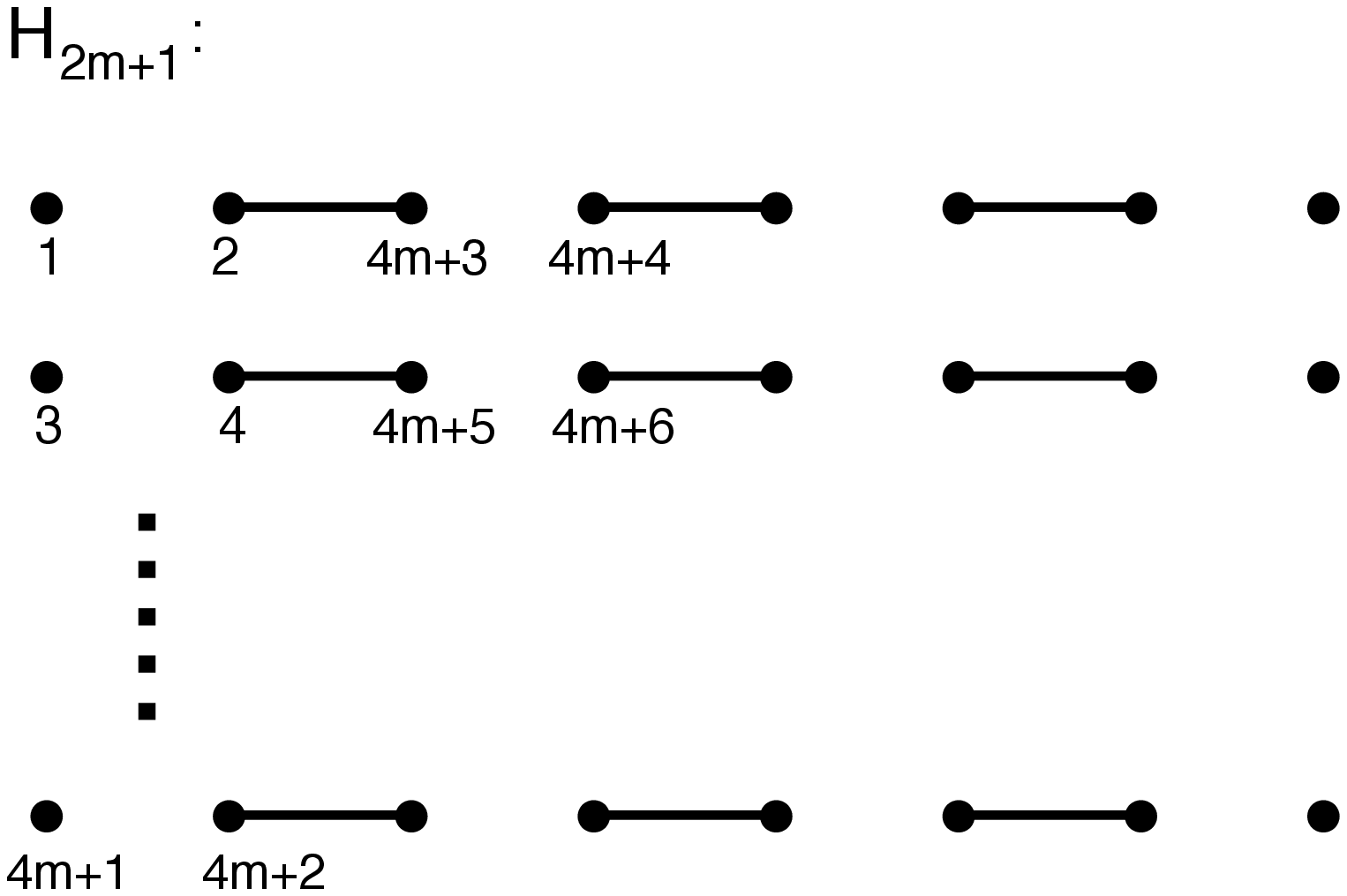}
\caption{Schematic representation of the Hamiltonian $H_{2m+1}$. \label{zfig4}}
\end{figure}

There are two alternatives. An \emph{even simple} algebra (over the field of complex numbers) has the form $\calA=\CC(\calL^{(0)}\oplus\calL^{(1)})$, where the even and odd parts are defined as follows:
\begin{equation}
\begin{pmatrix}X_{00}&0\\0&X_{11}\end{pmatrix}\in\calA^{(0)},\qquad
\begin{pmatrix}0&X_{01}\\X_{10}&0\end{pmatrix}\in\calA^{(1)}.
\end{equation}
Thus, the grading automorphism $\sigma_P$ is represented by an involutory operator $\hat{P}$:
\begin{equation}\label{hatPeven}
\sigma_{P}(X)=\hat{P}X\hat{P},\qquad \hat{P}
=\begin{pmatrix}\one_{\calL^{(0)}}&0\\[2pt]0&-\one_{\calL^{(1)}}\end{pmatrix}.
\end{equation}
The spaces $\calL^{(0)}$, $\calL^{(1)}$ consist of states with different local fermionic parity, though we cannot tell which is which since $\hat{P}$ is defined up to a sign. An \emph{odd simple} algebra does not have an internal grading operator, but rather an odd involutory central element  $\hat{Z}$.  For example, consider the model $H_{2m+1}$ in figure~\ref{zfig4} and think of the algebra $\calA_{l}$ of operators that act on the degenerate ground state and can be localized near the left end of the chain. This algebra is generated by  $k=2m+1$ unpaired Majorana modes, $c_1,c_3,\ldots,c_{4m+1}$ and has an odd involutory central element  $\hat{Z}_{l}=i^{m}c_1c_3\cdots c_{4m+1}$. In general, an odd simple algebra has the form $\calA=\CC(\calL)\oplus\CC(\calL)\kern1pt\hat{Z}$. Such algebras have a convenient representation on the Hilbert space $\CC^2\otimes\calL$: one sets $\hat{Z}$ to $M\otimes\one_{\calL}$, where $M$ is a suitable $2\times 2$ matrix. In the previous section we had $M=\sigma^z$, but it may be more natural to use $\sigma^y$ and $\sigma^x$ at the left and the right end of the chain, respectively. Thus, there are two standard forms of general algebra elements:  
\begin{equation}
\begin{pmatrix}X_0&0\\0&X_0\end{pmatrix}\in\calA^{(0)},\qquad
\begin{pmatrix}0&-iX_1\\iX_1&0\end{pmatrix}\in\calA^{(1)}
\label{oddrepl}
\end{equation}
or
\begin{equation}
\begin{pmatrix}Y_0&0\\0&Y_0\end{pmatrix}\in\calA^{(0)},\qquad
\begin{pmatrix}0&Y_1\\Y_1&0\end{pmatrix}\in\calA^{(1)}.
\label{oddrepr}
\end{equation}

For a system composed of pairs of Majorana modes, the left and right algebras are either both even or both odd. The full algebra $\calA_{\bound}$ is the graded tensor product $\calA_{l}\otimesgr\calA_{r}$, which implies a special multiplication rule. It is sufficient to define the multiplication for 
elements of the form $X\otimesgr 1$ and $1\otimesgr Y$, which we identify with $X$ and $Y$, respectively. If $X\in\calA_{l}^{(x)}$ and $Y\in\calA_{r}^{(y)}$\, ($x,y=0,1$), then
\begin{equation}
XY=X\otimesgr Y,\qquad YX=(-1)^{xy}\,X\otimesgr Y.
\end{equation}
In other words, operators at the two ends commute or anticommute according to their parity. This rule is natural if we think about physical fermions, but one can also construct such a product by modifying the usual tensor product, a trick appropriate in the Jordan-Wigner picture. In the even case, all operators act in the Hilbert space $\bigl(\calL^{(0)}_{l}\oplus\calL^{(1)}_{l}\bigr) \otimes \bigl(\calL^{(0)}_{r}\oplus\calL^{(1)}_{r}\bigr)$ with the obvious grading, and the (anti)commutation is achieved by defining $1\otimesgr Y$ as $\bigl(\hat{P}_{l}\bigr)^y\otimes Y$. In the odd case, the Hilbert space is $\calL_{\bound}=\CC^{2}\otimes\calL_{l}\otimes\calL_{r}$ with $\hat{P}=\sigma^{z}\otimes\one\otimes\one$, and the correct commutation relations are due to the use of representations~(\ref{oddrepl}), (\ref{oddrepr}).

Finally, we consider the action of symmetries on the simple $\ZZ_2$-graded algebra $\calA=\calA_{r}$. In the even case, $\calA=\CC(\calL^{(0)}\oplus\calL^{(1)})$ remains simple even if we ignore the grading. Therefore we can represent every automorphism $\sigma_g$ by a unitary or anti-unitary operator $\hat{g}$ (in particular $\sigma_P$ is represented by $\hat{P}$). This structure is just an extension of $G$ with $\UU(1)$ in the usual sense, but in the fermionic context we call it an \emph{even extension}. It automatically defines a homomorphism $\mu:G\to\ZZ_2$,
which tells whether a given symmetry $g$ preserves or changes the local fermionic parity:
\begin{equation}
\hat{g}\in\calA^{(\mu(g))},\qquad \mu(g)\in\{0,1\}.
\end{equation}
To see that $\hat{g}$ is indeed homogeneous, we write this condition as follows: $\hat{P}\hat{g}\hat{P}=(-1)^{\mu(g)}\hat{g}$, or
\begin{equation}\label{mu}
\sigma_{g}(\hat{P})=(-1)^{\mu(g)}\hat{P}.
\end{equation}
The algebra elements $\hat{P}$ and $\sigma_{g}(\hat{P})$ are related by a sign because they are both involutions and  represent the same automorphism, as shown by this calculation:
\[
\begin{aligned}
\sigma_{g}(\hat{P})\,X\,\sigma_{g}(\hat{P})
&=\sigma_{g}\bigl(\hat{P}\,\sigma_{g^{-1}}(X)\,\hat{P}\bigr)\\
&=\sigma_{g}\bigl(\sigma_P\bigl(\sigma_{g^{-1}}(X)\bigr)\bigr)
=\sigma_{P}(X).
\end{aligned}
\]
 It is clear that $\mu(P)=0$.

In the odd case, the algebra $\calA=\CC(\calL)\oplus\CC(\calL)\kern1pt\hat{P}$ is not simple without grading, but the subalgebra $\calA^{(0)}=\CC(\calL)$ is. Therefore the restriction of the automorphism $\sigma_g$ to $\calA^{(0)}$ can be represented by an operator  $\hat{g}'\in\calA^{(0)}$. Since $\sigma_P$ acts on $\calA^{(0)}$ trivially, we obtain an extension of the quotient group $G'=G/\{I,P\}$ with $\UU(1)$. To find the action of $\sigma_g$ on the whole algebra, we also need $\sigma_g(\hat{Z})$. It is defined by analogy with equation~(\ref{mu}), namely
\begin{equation}\label{mu1}
\sigma_{g}(\hat{Z})=(-1)^{\mu(g)}\hat{Z}.
\end{equation}
(This definition is correct because $\sigma_{g}$ preserves the properties of $\hat{Z}$ to be involutory, odd, and central, and such an operator is unique up to sign.) In the present case, the homomorphism $\mu$ is an independent piece of data satisfying the only constraint $\mu(P)=1$. For an arbitrary $X\in\calA_x$ we have
\begin{equation}
\sigma_g(X)=(-1)^{\mu(g)\,x}\, \hat{g}' X (\hat{g}')^{-1},
\end{equation}
where $\hat{g}'$ is constant on each coset $\{g,gP\}$.

One may wonder why the structure of even and odd extensions is so different. We cannot give an exact answer but notice that odd extensions do not exist for all groups. Indeed, a homomorphism $\mu:G\to\ZZ_2$ with the property $\mu(P)=1$ defines an embedding $\tau:G'\to G$. (In each coset $\{g,gP\}$ we pick the element $\tau(g)=h$ such that $\mu(h)=0$.) Thus, $G\cong G'\times\{I,P\}$. In particular, $P$ cannot have a square root in $G$. This precludes odd extensions for the groups $G=\ZZ_{4m}$ and their limiting case $\UU(1)$, which is consistent with the fact that unpaired Majorana modes only occur in superconducting systems.

On the other hand, if $G=G'\times\{I,P\}$, then even and odd extensions can be unified. They are both defined by an (ordinary) extension of $G'$ with $U(1)$ and a homomorphism $\mu:G\to\ZZ_2$. The latter can be specified by the value of $\mu(P)$ and a homomorphism $\mu':G'\to\ZZ_2$. For the TR-invariant Majorana chain, we have $G'=\{I,T\}$ where $T$ is anti-unitary. Thus, a general extension is given by the invariants $\epsilon=(-1)^{\mu(P)}$,\: $a=(-1)^{\mu(T)}$, and $\hat{T}^2=\pm1$. The latter may be interpreted as a choice between real numbers ($D=\RR$) and quaternions ($D=\HH$). These are the Wall invariants in table~\ref{table1}. 

\section{Calculation of the invariants \label{ex}}

We have defined interacting invariants which distinguish eight possible phases of the Majorana chain.  We now show, for completeness, that all eight phases are realized in the band TR-invariant Majorana chain, and relate the band topological index $k$ to the values of the interacting invariants from the previous section.  Here we essentially re-derive the mathematical structure of the previous section, for the specific case of the TR-invariant Majorana chain, in a much more pedestrian fashion.

We consider the flat band Majorana chain Hamiltonians $H_k$ defined in figure \ref{zfig3}. We need to compute $\hat{P}$ and $\hat{T}$ on ${\cal H}_L$, the Hilbert space spanned by the Schmidt eigenstates of the left half of the chain.  (It is similar to the space of low energy states $\calL_r$ that was used in the previous section.)  We begin by taking $k$ even.

\subsection{Even $k$}

Let $k=2m$.  The picture is as in figure \ref{zfig3}, where we take the dashed vertical line to represent the right endpoint of the system.  Then the space of low energy operators $\CC(\H_L)$ is spanned by the unpaired Majorana modes $c_2, c_4, \ldots, c_{2k}$.  First, consider the action of the parity operator $P$, defined as the product of physical fermion occupation numbers over all sites.  Because the bulk Majorana modes are paired into unoccupied physical fermions, $P$ reduces at low energy to the parity operator for the $2m$ Majoranas $c_2, \ldots, c_{2k}$:  \begin{equation} \label{projp} \hat{P} = (i c_2 c_4) \cdots (i c_{4m-2} c_{4m}).\end{equation} 

Second, consider the action of $\hat{T}$.  By definition, $\hat{T} c_{2l}\, \hat{T}^{-1} = c_{2l}$ for $l \in {1, \ldots, 2m}$, and $\hat{T}\, i\, \hat{T}^{-1} = -i$. Thus from (\ref{projp}) we see that
 \begin{equation}\label{a_even}
\hat{T} \hat{P}\kern1pt \hat{T} ^{-1} = a\hat{P},\qquad a=(-1)^m.
\end{equation}

Now we need to determine the sign of $\hat{T}^2$.  In general, for a $2^j$ dimensional Hilbert space ${\cal H}'$ of the Majoranas $c_1', \ldots, c_{2j}'$, the action of an anti-unitary $T'$ is uniquely determined up to phase by its adjoint action on the $c_l'$.  Indeed, given two anti-unitaries $T'$ and $T''$ with the same adjoint action, the unitary  $(T')^{-1} T''$ commutes with the entire Clifford algebra, so that $T''$ is equal to $T'$  up to a phase. In particular, because $T'$ is anti-unitary, the sign of $(T')^2$ is uniquely determined.  Now, we know that for even indices $2l$, \begin{equation} \label{cr} \hat{T} c_{2l} \hat{T}^{-1} = c_{2l} \end{equation} for $l \in {1, \ldots, 2m}$.  Let us find an explicit expression for $\hat{T}$ in the occupation number basis, assuming that the annihilation and creation operators $b_r, b_r^\dag$\: ($r =1, \ldots, m$) are defined as follows:  \begin{eqnarray} c_{4r-2} &=& -i (b_r - b_r^\dag) \nonumber\\ c_{4r} &=& b_r + b_r^\dag. \end{eqnarray} Let  $|0\rangle$  be the state annihilated by all the $b_r$.  Any $|\psi\rangle \in {\cal H}_L$ can be written as  \begin{equation} |\psi\rangle = \sum_{\alpha_i \in \{0,1\}} C_{\alpha_1 \ldots \alpha_m} {b_1^\dag}^{\alpha_1} \cdots {b_m^\dag}^{\alpha_m} |0\rangle. \end{equation}  Let $K$ be the complex conjugation operator on this space of wavefunctions: that is, $K$ acts by complex conjugating the coefficients $C_{\alpha_1 \ldots \alpha_m}$.  Now define
\begin{equation} \label{t12}
\hat{T} =\left\{ \begin{array}{ll}
c_2 c_6 \cdots c_{4m-2} K & \text{for $m$ even}, \\
c_4 c_8 \cdots c_{4m} K & \text{for $m$ odd}.
\end{array}\right.
\end{equation}
We check that $\hat{T}$, so defined, satisfies (\ref{cr}), and hence is the unique choice, up to sign, of the projective action of $T$ on $\calH_L$.  We now find by direct computation that $\hat{T}^2 = 1$ for $m\equiv 0,1$ modulo $4$, and $\hat{T}^2 = -1$ for $m \equiv 2,3$ modulo $4$, so that $\hat{T}^2 = 1$ for $k=0,2$, and $\hat{T}^2 = -1$ for $k=4,6$.

Thus we see that the sign of the commutation relation between $\hat{P}$ and $\hat{T}$, and the sign of $\hat{T}^2$, together uniquely determine even $k$ modulo $8$.  We gather these results in table~\ref{table1}.

\subsection{Odd k}

Now let $k=2m+1$.  We proceed as in the previous subsection, but the situation is now complicated by the two degenerate ground states.  We need to construct the Hilbert space ${\cal H}_L$ spanned by the Schmidt eigenstates of both of these ground states.  The setup again is as in figure \ref{zfig3}, this time with an odd number of Majoranas, $c_2, \ldots, c_{4m+2}$, lacking partners.  This situation may seem problematic, as we need an even number of Majoranas to define a physical Hilbert space.  It is remedied by the the existence of an extra Majorana mode $c_\infty$ at infinity, which appears whenever we try to cut the system off at a finite size.  Thus ${\cal H}_L$ is the $2^{m+1}$ dimensional Hilbert space of the Majoranas $c_2, \ldots, c_{4m+2}, c_\infty$.

As before, the parity operator is just the product \begin{equation} \label{defp} \hat{P} = (i c_2 c_4) \cdots (i c_{4m-2} c_{4m}) (i c_{4m+2} c_\infty). \end{equation}  It is also useful to introduce an operator $\hat{Z}$ that is similar to $\hat{P}$ but does not involve $c_\infty$:
\begin{equation}
\hat{Z}=i^{m}c_2\cdots c_{4m+2}.
\end{equation}
It squares to $1$, commutes with all operators acting locally (i.e.\ with combinations of  $c_2,\ldots,c_{4m+2}$), and anticommutes with $\hat{P}$. Under the time reversal symmetry, $\hat{Z}$ transforms like $\hat{P}$ in the even case:
\begin{equation}\label{a_odd}
\hat{T} \hat{Z}\kern1pt \hat{T} ^{-1} = a\hat{Z},\qquad a=(-1)^m.
\end{equation}
However,  we now have an ambiguity, because we have not specified the transformation of $c_\infty$ under $\hat{T}$.  In fact, there are two possible choices for a consistent anti-unitary symmetry, $\hat{T} c_\infty \hat{T}^{-1} = \pm c_\infty$.  By analogy with the MPS construction in section~\ref{ii}, we choose the sign so that $\hat{T}$ and $\hat{P}$ commute, namely
\begin{equation}
\hat{T} c_\infty \hat{T}^{-1} = (-1)^{m+1} c_\infty.
\end{equation}
Similarly to (\ref{t12}), we now pair up the $c_j$ into annihilation and creation operators as indicated in the expression (\ref{defp}) for $\hat{P}$, and in this basis define
\begin{equation}
\hat{T} =\left\{ \begin{array}{ll}
c_2 c_6 \cdots c_{4m+2} K & \text{ for $m$ odd} \\
c_2 c_6 \cdots c_{4m+2} c_\infty K & \text{ for $m$ even}.
\end{array}\right.
\end{equation}
(In the second case, one can represent $\hat{T}$ as $\hat{T}'\hat{P}$, where $\hat{T}'$ does not involve $c_\infty$. Although $\hat{T}'$ does not implement the time reversal symmetry on the Majorana operators, it acts correctly on physical observables, i.e.\ even products of $c_2,\ldots,c_{4m+2}$).

A simple calculation shows that $\hat{T}^2 =(\hat{T}')^2 = -1$ for $m\equiv 1,2$ modulo $4$, and $\hat{T}^2 =(\hat{T}')^2 =1$ for $m\equiv 0,3$ modulo $4$. Thus, $\hat{T}^2=1$ for $k=1,7$ and $\hat{T}^2=-1$ for $k=3,5$. The results are gathered in table~\ref{table1}.

\begin{table*} 
\extrarowheight=2pt
\begin{tabular}{|c||c|c|c|c||c||c|c|c|c|}
\hline
topological index & \multicolumn{4}{c||}{Wall invariants} &
& \multicolumn{4}{c|}{Altland-Zirnbauer classification}\\
\hhline{|~||-|-|-|-||~||-|-|-|-|}
$k\bmod 8$ & $\kk$ & $\epsilon$ & $a$ & $D$ & $\hat{T}^2$
& $\hat{T}_{+}^2$ & $\hat{T}_{-}^2$ & $\hat{Z}^2$ & Cartan label\\
\hhline{|=#=|=|=|=#=#=|=|=|=|}
$0$ & $\RR$ & $+$ & $+1$ & $\RR$ & $+1$
 & $+1$ &      &     & AI (orthogonal) \\ \hline
$1$ & $\RR$ & $-$ & $+1$ & $\RR$ & $+1$
 & $+1$ & $+1$ & $1$ & BDI (chiral orthogonal) \\ \hline
$2$ & $\RR$ & $+$ & $-1$ & $\RR$ & $+1$
 &      & $+1$ &     & D (BdG) \\ \hline
$3$ & $\RR$ & $-$ & $-1$ & $\HH$ & $-1$
 & $-1$ & $+1$ & $1$ & DIII (BdG) \\ \hline
$4$ & $\RR$ & $+$ & $+1$ & $\HH$ & $-1$
 & $-1$ &      &     & AII (symplectic) \\\hline
$5$ & $\RR$ & $-$ & $+1$ & $\HH$ & $-1$
 & $-1$ & $-1$ & $1$ & CII (chiral symplectic) \\ \hline
$6$ & $\RR$ & $+$ & $-1$ & $\HH$ & $-1$
 &      & $-1$ &     & C (BdG) \\ \hline
$7$ & $\RR$ & $-$ & $-1$ & $\RR$ & $+1$
 & $+1$ & $-1$ & $1$ & CI (BdG) \\
\hhline{|=#=|=|=|=#=#=|=|=|=|}
    & $\CC$ & $+$ & $+1$ & $\CC$ &
 &      &      &     & A (unitary) \\ \hline
    & $\CC$ & $-$ & $+1$ & $\CC$ &
 &      &      & $1$ & AIII (chiral unitary) \\ \hline
\end{tabular}
\caption{The correspondence between the topological index $k$ of a TR-invariant Majorana chain, the Wall invariants of simple $\ZZ_2$-graded algebras~\cite{Wall}, and the Altland-Zirnbauer classes of free-fermion Hamiltonians~\cite{AZ,Zirnbauer,HHZ} (using the scheme~\cite{SRFL,RSFL1}). Both classifications are explained in appendix~\ref{AZW}, and the invariants $a$ and $D$ for the Majorana chain are calculated in section~\ref{ex}.}
\label{table1}
 \end{table*}

\section{Discussion \label{d}}
We have constructed interacting physical invariants which distinguish among the eight $T$ protected phases of the TR-invariant Majorana chain, and related them to Altland-Zirnbauer symmetry classes of matrices.  This TR-invariant Majorana chain is in some sense the simplest one dimensional system incorporating both fermions and anti-unitary symmetries.  Its solution led us to a general framework for classifying gapped phases of such chains with an arbitrary symmetry group $G$, that of  extensions of $G$ with $\UU(1)$. 

For the case of the TR-invariant Majorana chain, one can tell which phase one is in by looking at the ground state reduced density density matrix $\rho$ of a semi-infinite chain.  The key is to study the symmetries of $\rho$ considered as an operator on the entanglement Hilbert space, which is defined as the span of the, say, left Schmidt eigenstates: $\rho: \calH_L \rightarrow \calH_L$.  $\calH$ carries additional structure coming from a projective representation $\hat{P}$, $\hat{T}$ of the underlying symmetries.  In addition, when one has unpaired Majorana modes there is a symmetry $\hat{Z}$ which distinguishes the ground states.  This structure amounts to a specification of Altland-Zirnbauer symmetry class for $\rho$; one aspect of this interpretation is that the anti-unitary symmetry exchanging different fermionic parity ($\hat{P}$) sectors is thought of as  a particle-hole symmetry ($\hat{T}_{-}$).

The parity of $k$ has a familiar physical interpretation: it describes the presence or absence of unpaired Majorana modes at the endpoints of a sample.  What about the invariants which determine $k$ modulo $4$ and $8$?  In certain limits, these invariants reduce to other, more well known ones.  For example, consider restricting to insulators, by disallowing pairing terms in the Hamiltonian.  This eliminates the four superconducting phases with odd $k$, but still leaves four distinct phases corresponding to even $k=2m$.  Then even and odd $m$ correspond respectively to the two phases of polyacetelene.  Indeed, it is well known that a domain wall between these two phases supports a gapless mode, and carries fractional charge $\pm 1/2$ depending on the occupation of this mode \cite{Sus}.  Now, we can view $\hat{P}$ and $\hat{T}$ as effectively acting on this gapless mode, and when they anti-commute, i.e. when $m$ is odd, there is a Kramers pair of edge states with opposite fermionic parity.  In the simplest case, by symmetry, they must have charge $\pm 1/2$.  Hence the invariant that determines even $k$ modulo $4$ recovers the well-known physics of the Su-Schrieffer model.  This invariant also has a field theory interpretation, as the presence or absence of a non-trivial $\theta=\pi$ theta term in the effective $U(1)$ gauge action for the theory.

What about $k$ modulo $8$?  In trying to distinguish between the phase of Hamiltonian $H_4$ and the trivial phase, we can impose the $SO(4)$ symmetry, which rotates the four chains into one another.  The non-trivial phase of $H_4$ has edge modes that transform as $(1/2,0) \oplus (0,1/2)$ under $so(4) = su(2) \oplus su(2)$.  The generic Hamiltonian of this phase will split the degeneracy between these spin $1/2$'s, but will retain at least one gapless spin $1/2$.  Hence the non-trivial and trivial phases are distinguished by the presence or absence of a spin $1/2$ edge mode.  This is reminiscent of the $AKLT$ state, and indeed, in the space of $su(2) \subset so(4)$ symmetric Hamiltonians the non-trivial phase represented by $H_4$ is precisely the Haldane phase.

Thus the invariant $k$ modulo $8$ ties together and generalizes several well known one dimensional physical invariants.  It would be interesting to generalize these methods to higher dimensions.

\acknowledgments We would like to acknowledge useful discussions with Jason Alicea, Matthew Hastings, Netanel Lindner, John Preskill, Gil Refael, Ari Turner, and Dan Freed.  A.K. is grateful to the Aspen Center for Physics for hospitality.  This work was supported in part by the institute for Quantum Information under National Science Foundation grant no.\ PHY-0803371.

\appendix

\section{Altland-Zirnbauer and Wall classes \label{AZW}}

 In this appendix, we reveal the exact mathematical structure that is common to the TR-invariant Majorana chain and the Altland-Zirnbauer theory. The AZ classification is concerned with a different set of physical systems, free-fermion Hamiltonians  (in dimension $0$). In this setting, the symmetry group $G$ acts linearly on $a_j,a^\dag_j$, or equivalently, on $c_j$.  As is usual, the group elements are marked as ``unitary'' or ``anti-unitary'' (using a homomorphism $\alpha:G\to\ZZ_2$). Because both types of  symmetries preserve Hermicity, we have a linear action of $G$ in the \emph{mode space} $\calM=\RR^{2N}$  that consists of  operators of the form $X=\sum_lx_lc_l$\: ($x_l\in\RR$). The Majorana modes $c_l$ form an orthonormal basis of $\calM$, and the inner product between two arbitrary elements is given by the anticommutator (more exactly, $\frac{1}{2}\{X,Y\}$). Thus, each symmetry $g\in G$ is represented by a real orthogonal matrix $S_g\in\OO(2N)$. We are interested in characterizing all quadratic Hamiltonians~(\ref{quadH}) that are invariant under this action. This invariance translates to the following condition on the matrix $A$ in~(\ref{quadH}):
\begin{equation}\label{Ainv}
S_gAS_g^{-1} = \left\{ \begin{array}{ll}
A, &\text{if $g$ is unitary},\\[2pt]
-A, &\text{if $g$ is anti-unitary}.
\end{array}\right.
\end{equation}

This problem was formulated and solved in \cite{HHZ} using the Nambu space language.  (A transparent, albeit brief exposition can be found in~\cite{RSFL1}.)  The Nambu space consists of complex linear combinations of $a_j,a^\dag$, or equivalently, of $c_l$. Therefore the Nambu space $\calN=\CC^{2N}$ is just the complexified mode space. Conversely, the mode space $\calM\subset\calN$ is the real subspace fixed by the conjugation $X\mapsto X^\dag$, which is an anti-unitary transformation of the Nambu space. In the conventional basis of $a_j,a^\dag_j$, it is written as $\hat{C}=U_{C}K$. To extend the invariance condition to the Nambu space one could simply use the same  equation~(\ref{Ainv}), but tradition requires that the equations look similar to those in many-body quantum mechanics. So let us introduce a ``single-particle Hamiltonian'' $\Hsp=\frac{i}{4}A$ (not bothering with basis changes) and redefine the symmetries, so that for anti-unitary $\hat{g}$, the corresponding operator on the Nambu space is likewise anti-unitary:
\begin{equation}
\hat{g} = \left\{ \begin{array}{ll}
S_g, &\text{if $g$ is unitary},\\[2pt]
S_g\hat{C}, &\text{if $g$ is anti-unitary}.
\end{array}\right.
\end{equation}
Note that the $\hat{g}$ commute with $\hat{C}$. Now, the Hamiltonian satisfies these conditions:
\begin{eqnarray}
\hat{g}\,\Hsp\,\hat{g}^{-1} &=& \Hsp\quad \text{for all } g\in G,
\nonumber\\[2pt]
\hat{C}\,\Hsp\,\hat{C}^{-1} &=& -\Hsp.
\end{eqnarray}

The solution turns out to be rather insensitive to the symmetry group $G$. Indeed, suppose first that $G$ consists of only unitary symmetries. The action of $G$ then defines a \emph{block decomposition},
\begin{equation}\label{blocks}
\calN=\bigoplus_{\lambda}\calL_\lambda\otimes\calH_\lambda,
\end{equation}
where $\lambda$ indexes the irreps $\calL_\lambda$, and the Hamiltonian acts independently in each $\calH_\lambda$ (so that $\dim \calH_\lambda$ is the multiplicity of $\calL_\lambda$). The conjugation discriminates between real irreps, quaternionic irreps, and conjugate pairs of complex irreps; each choice corresponds to a certain form of possible Hamiltonians for the given block. For example, suppose that there is only one block $\calL\otimes\calH$ of quaternionic type, meaning that $\hat{C}_{\calL}^{2}=-1$. But the conjugation on the total space satisfies the condition $\hat{C}^2=1$ and factors as $\hat{C}=\hat{C}_{\calL}\otimes\hat{C}_{\calH}$, hence $\hat{C}_{\calH}^2=-1$. The anticommutation between $\hat{C}_{\calH}$ and the Hamiltonian implies that $\Hsp\in\spl(n)$ (the Cartan class $C$). Similarly, real blocks and conjugate pairs of complex blocks yield Hamiltonians in the $D$ and $A$ Cartan classes, respectively.

When physical anti-unitary symmetries are allowed, the number of cases increases to $10$. The set of possible Hamiltonians (within a single block) is defined by commutation relations with a few operators that remain after factoring out unitary symmetries. Such operators may be \emph{even} or \emph{odd}, depending on whether they commute or anticommute with the Hamiltonian.  Up to three special operators may be present, which are labeled as follows:
\begin{itemize}
\item $\hat{T}_{+}$: even anti-unitary,\, $\hat{T}_{+}^2=\pm1$ (TRS);
\item $\hat{T}_{-}$: odd anti-unitary,\, $\hat{T}_{-}^2=\pm1$ (PHS);
\item $\hat{Z}$: odd unitary,\, $\hat{Z}^2=1$ (SLS).
\end{itemize}
(The abbreviations stand for ``time-reversal symmetry'', ``particle-hole symmetry'', and ``sublattice symmetry'', though they are not concrete elements of the symmetry group.) Furthermore, if two of these operators exist, the third is defined by the identity
\begin{equation}
\hat{Z}=\hat{T}_{+}\hat{T}_{-}.
\end{equation}
All $10$ cases are listed in table~\ref{table1}. Note that in this scheme, the special operator $\hat{C}$ mixes with physical symmetries and may disappear in the unitary elimination process.

A mathematically more appealing way to arrive at the same result is to work with real coefficients\footnote{ Here we follow Dyson's philosophy \cite{Dyson}. In fact, he also found ten symmetry classes, though a coarser division into only 3 classes appears in the final answer. This is because Dyson considers the action of symmetries on the Hilbert space of a many-body system. Unlike the Nambu space, the Hilbert space of quantum states has no canonical conjugation, since the multiplication by a phase is an intrinsic symmetry. } and use equation~(\ref{Ainv}) directly. The technical trick,
\[
\text{real space}\:\leftrightarrow\:
\text{complex space with a conjugation}
\]
may be used at some point, but the conjugation will always be identified as such.  The parity of operators is defined according to their commutation with $A$; in particular, the $S_g$ are even for unitary $g$ and odd for anti-unitary $g$. We now replace the symmetry  group by the algebra $\calA=\calA^{(0)}\oplus\calA^{(1)}$, where $\calA^{(0)}$ and $\calA^{(1)}$ consist of formal linear combinations of unitary and anti-unitary elements $g\in G$, respectively.  We call $\calA^{(0)}$ and $\calA^{(1)}$ the \emph{even} and \emph{odd} parts of $\calA$. Thus, $\calA$ is a semisimple associative $\ZZ_2$-graded algebra over the field of reals, and its action on $\calM$ encodes all relevant information about the symmetries.  (The simplicity and semisimplicity of graded algebras are defined in terms of representations as in the ungraded case, see appendix~\ref{semsimp}. However, intertwiners can be \emph{even} or \emph{odd}; the odd ones commute with $\calA^{(0)}$ but anticommute with $\calA^{(1)}$.) 

The solution to this more abstract problem also uses a block decomposition. Although the blocks of the mode space do not factor as $\calL_\lambda\otimes\calH_\lambda$, they carry two simple graded algebras: the symmetry algebra $\calA_\lambda$ (a simple component of $\calA$) and the algebra $\calB_\lambda$ of operators that (anti)commute with these symmetries. The ``Hamiltonian'' for each block belongs to $\calB_\lambda^{(1)}$, the odd part of $\calB_\lambda$. This method will be presented elsewhere. It is relatively straightforward, thanks to Wall's complete theory of simple $\ZZ_2$-graded algebras (over an arbitrary field) \cite{Wall}.

We now give a short summary of Wall's invariants in the real case. We first reproduce the original definition (mainly, for completeness). Then we describe it in terms of complex algebras with conjugation, the language most natural in the Majorana chain context. (See section~\ref{mc} for the classification of complex $\ZZ_2$-graded algebras with general symmetries.) Finally, we formulate a variant that is most closely related to the AZ scheme, see equations (\ref{commrel1}), (\ref{commrel2}). 

Let $\calA=\calA^{(0)}\oplus\calA^{(1)}$ be a simple $\ZZ_2$-graded algebra over the reals. Four invariants are defined successively.
\begin{enumerate}
\item \emph{Graded center} $\kk=Z(\calA)\cap \calA^{(0)}$, where $Z(\calA)$ is the center of $\calA$ in the usual sense. $\kk$ is a field, i.e.\ $\RR$ or $\CC$. It is convenient to consider $\calA$ as a graded algebra over $\kk$. The invariants $a$ and $D$ (see below) are trivial for complex algebras.

\item A label $\epsilon$, either $+$ or $-$. In both cases, there is an ungraded simple algebra $\calD$ with center $\kk$ and an invertible element $u\in\calA$, though they have somewhat different properties.
\begin{itemize}
\item[($+$)] \emph{(even case):}\: $\calD=\calA$ (as an ungraded algebra); $u$ belongs to $\calA^{(0)}$ and defines the grading, i.e. $uXu^{-1}=(-1)^{x}X$ for any $X\in\calA^{(x)}$\: ($x=0,1$).
\item[($-$)] \emph{(odd case):}\: $\calD=\calA^{(0)}$, whereas $u$ belongs to $\calA^{(1)}$ and commutes with all $X\in\calA$. It follows that $\calA^{(1)}=u\calD$.
\end{itemize}

\item A nonzero number $a=u^2\in\kk$. (It is clear that $u^2$ belongs to $\calA^{(0)}$ and commutes with everything, hence it is in the graded center.) Because $a$ is defined up to multiplication by a square, we may assume that $a=\pm1$ if $\kk=\RR$ and $a=1$ if $\kk=\CC$. 

\item The type $D$ of the ungraded simple algebra $\calD$: real or quaternionic if $\kk=\RR$, complex if $\kk=\CC$.
\end{enumerate}
We can readily see that there are $10$ possibilities: $8$ with $\kk=\RR$ and $2$ with $\kk=\CC$ (see table~\ref{table1}). 

Let us now specialize to $\kk=\RR$ (the case relevant to the Majorana chain) and represent the above structure by commutation relations between some operators. To this end, we replace $\calA$ with its complexified version $\calA_\CC$ (by adjoining the imaginary unit). One can prove that $\calA_\CC$ is simple as a $\ZZ_2$-graded algebra. Multiplying the special element $u$ by a suitable complex number, we obtain $\hat{W}\in\calA_\CC$ such that $\hat{W}^2=1$. In the main part of the paper, we denoted $\hat{W}$ by either $\hat{P}$ or $\hat{Z}$:
\begin{equation}
\hat{W}=\left\{ \begin{array}{ll}
\hat{P} &\text{in the even case},\\[2pt]
\hat{Z} &\text{in the odd case}.
\end{array}\right.
\end{equation}
The invariant $a$ is defined by these equations:
\begin{equation}\label{commrel}
\hat{W}^{2}=1,\qquad \sigma_{T}(\hat{W})=a\hat{W},
\end{equation}
where $\sigma_T$ is the complex conjugation on $\calA_\CC$.

In the even case, $\calA_\CC$ is isomorphic to $\CC(\calL)$ as an ungraded algebra, and $\sigma_T$ can be represented by an anti-unitary operator $\hat{T}$ acting in $\calL$. In the odd case, this does not work directly because $\calA_\CC\cong\CC(\calL)\oplus\CC(\calL)\kern1pt\hat{Z}$ is not simple as an ungraded algebra. To achieve the desired representation, we double the space $\calL$, set $\hat{W}=\hat{Z}=\sigma^z\otimes\one_\calL$, and introduce an auxiliary operator $\hat{P}=\sigma^x\otimes\one_\calL$. (This procedure is described more concretely in section~\ref{ii}.) Now we can represent $\sigma_T$ by $\hat{T}=\one\otimes\hat{T}'$ if $a=1$, or by $\hat{T}=\sigma^x\otimes\hat{T}'$ if $a=-1$. The whole construction is characterized by the following equations.
\begin{align}
&\text{Even case:} &&\hat{P}^{2}=1, &&\hat{T}\hat{P}=a\hat{P}\hat{T}; \label{commrel1}\\[5pt]
&\text{Odd case:} && \hat{Z}^{2}=1, && \hat{T}\hat{Z}=a\hat{Z}\hat{T},
\label{commrel2}\\[2pt]
&&& \hat{P}^2=1, && \hat{P}\hat{T}=\hat{T}\hat{P},\quad
\hat{P}\hat{Z}=-\hat{Z}\hat{P}.
\nonumber
\end{align}
 In both cases, $\hat{T}^2=\pm1$ defines the ungraded type, i.e.\ $\RR$ or $\HH$.

To make a correspondence with the AZ scheme, notice that $\hat{T}$ can be labeled as $\hat{T}_{+}$ or $\hat{T}_{-}$, depending on its commutation with $\hat{P}$. In the even case, this depends on $a$. In the odd case, we have $\hat{T}_{+}=\hat{T}$, while $\hat{T}_{-}$ is defined as $\hat{T}^{-1}\hat{Z}$. It follows that $\hat{T}_{-}^{2}=a\hat{T}^{2}$.   All cases are listed in table~\ref{table1}.\footnote{ Note a possibility of confusion: the two methods produce different values of AZ invariants for the same block of a free-fermion Hamiltonian because the operator parity is defined by the commutation with $A$ in one case and $\Hsp$ in the other. This discrepancy can be fixed by swapping $\hat{T}_{+}$ and $\hat{T}_{-}$, or by changing $k$ to $2-k$. We have chosen not to do that because the variable change $k\mapsto 2-k$ is convenient in the free-fermion context, see~\cite{kitaev-2009}. }

\section{Semisimple algebras \label{semsimp}}

For the reader's convenience, we summarize some basic facts about semisimple finite-dimensional associative algebras over $\RR$ or $\CC$. This is quite standard material, see e.g.~\cite{FarbDennis}. An (associative) algebra $\calA$ over a field $\kk$ is an associative ring (i.e., it has addition and multiplication with the usual properties) that contains $\kk$ in its center. Here are some examples:
\begin{enumerate}
\item $\kk(n)$, the algebra of all $n\times n$ matrices whose entries are elements of the base field $\kk$. It is the same as the algebra $\kk(\calL)$ of linear operators acting in the space $\calL=\kk^n$.
\item The real numbers ($\RR$), complex numbers ($\CC$), and quaternions ($\HH$) regarded as algebras over $\RR$. These are, in fact, \emph{division algebras}, i.e.\ all nonzero elements are invertible. According to the Frobenius theorem, there are no other real division algebras. The only complex division algebra is $\CC$.
\item The \emph{group algebra} $\kk[G]$ of a finite group $G$. It consists of linear combinations $x=\sum_{g\in G}x_g\kern1pt\mathbf{e}_g$, where $x_g\in\kk$, and the basis elements $\mathbf{e}_g$ satisfy the relations $\mathbf{e}_g\mathbf{e}_h=\mathbf{e}_{gh}$.
\item The algebra generated by $\mathbf{1}$ and $\mathbf{e}$, where $\mathbf{e}^2=0$. (A general element has the form $x+y\kern1pt\mathbf{e}$, where $x,y\in\kk$.)
\end{enumerate}
In this list, items~1 and~2 are simple algebras, 3 is semisimple (for $\kk=\RR,\CC$) but generally not simple, and 4 is not semisimple.

The simplicity and semisimplicity are defined in terms of (finite-dimensional) \emph{representations}, i.e. vector spaces over the base field on which the algebra acts linearly. One also needs the notion of \emph{intertwinwer}: a linear map between two representations that commutes with the algebra action. An \emph{isomorphism} is an invertible intertwiner.

An algebra $\calA$ is called \emph{semisimple} if any subrepresentation $\calL\subseteq\calN$ has a complementary representation $\calM$, i.e.\ $\calN=\calL\oplus\calM$. (This property holds if $\calA$ is a matrix subalgebra that is closed under the Hermitian conjugation, one example being the group algebra represented by permutation matrices. Indeed, in this case any representation $\calN$ is \emph{unitary}, i.e.\ it has a Hermitian inner product such that conjugate elements $X,X^\dag\in\calA$ are represented by adjoint operators. Thus, the orthogonal complement $\calM=\calL^{\perp}$ is invariant under the algebra action.) Any representation of a semisimple algebra splits into irreducible ones. A semisimple algebra is called \emph{simple} if it has a unique (up to isomorphism) irreducible representation.

A theorem of Wedderburn states that any semisimple algebra is a direct sum of simple algebras, and any simple algebra is isomorphic to the algebra of matrices $D(n)\cong D\otimes\kk(n)$, where $D$ is a division algebra. $D$ defines a \emph{type} (or \emph{equivalence class}) of simple algebras. For example, the group algebra $\calA=\CC[G]$ splits as $\calA\cong\bigoplus_\lambda\CC(\calL_\lambda)$, where $\calL_\lambda$ are the irreps of $G$. In the real case, a similar decomposition involves real, quaternionic, and complex matrices.

It is often useful to consider the tensor product of two simple algebras over the base field. This is the ``multiplication table'' for real division algebras:
\begin{eqnarray}
\CC\otimes_{\RR}\CC &\cong& \CC\oplus\CC,\\
\HH\otimes_{\RR}\CC &\cong& \CC(2),\\
\HH\otimes_{\RR}\HH &\cong& \RR(4).
\end{eqnarray}
Note that the last two products are simple. That is a special case of the following theorem (which also holds for base fields other than $\RR$ and $\CC$ and can be generalized to $\ZZ_2$-graded algebras): \emph{If $\calA$, $\calB$ are simple algebras over $\kk$, and the center of $\calA$ is equal to $\kk$, then $\calA\otimes_{\kk}\calB$ is simple}.

Let $\calA$ be a simple algebra with center $\RR$. While the Wedderburn and Frobenius theorems imply that $\calA$ is isomorphic to a real or quaternionic matrix algebra, there is an independent way to characterize $\calA$. (This argument can actually be used to prove the Frobenius theorem.) Consider the complex algebra $\calA_\CC=\calA\otimes_\RR\CC$.  It is simple, and hence isomorphic to $\CC(n)$. Thus, the complex conjugation $\sigma_T:\calA_\CC\to\calA_\CC$ can be implemented by some antilinear operator $\hat{T}$, meaning that
\begin{equation}
\sigma_T(X)=\hat{T}X\hat{T}^{-1}\quad \text{for all } X\in\calA_\CC.
\end{equation}
We may write $\hat{T}=U_{T}K$, where $U_{T}$ is an $n\times n$ matrix and $K$ is the complex conjugation on $\CC^{n}$. It is easy to see that $\hat{T}^2$ is a nonzero real number. (If the representation $\calL$ is unitary, we may assume that $\hat{T}$ is anti-unitary so that $\hat{T}^2=\pm1$.) Now, if $\hat{T}^2>0$, then $\calA\cong\RR(n)$; otherwise $\calA\cong\HH(n/2)$.

\bibliographystyle{aipproc-ctitle}

\vspace{5mm} 
\bibliography{z8refs}

\end{document}